\documentclass[conference]{IEEEtran}
\IEEEoverridecommandlockouts

\usepackage{amsmath,amssymb,amsthm,mathrsfs,amsfonts,dsfont} 
\DeclareMathAlphabet{\mathpzc}{OT1}{pzc}{m}{it}
\usepackage{graphicx}
\usepackage{textcomp}
\usepackage{algorithm,algorithmic,setspace}
\usepackage{tabularx,booktabs}
\usepackage{adjustbox}
\usepackage{array,etoolbox}
\usepackage{dcolumn}
\usepackage{rotating}
\usepackage{pifont}
\usepackage{tikz}
\usepackage{xcolor}
\usepackage{comment}
\usepackage{dsfont}
\usepackage{multicol}

\usepackage[nopostdot,style=super,nonumberlist, toc]{glossaries}
\usepackage{tabularx,booktabs}
\usepackage[utf8]{inputenc}
\usepackage{fourier} 
\usepackage{array}
\usepackage{makecell}
\newcolumntype{C}{>{\centering\arraybackslash}X} 
\def\BibTeX{{\rm B\kern-.05em{\sc i\kern-.025em b}\kern-.08em
    T\kern-.1667em\lower.7ex\hbox{E}\kern-.125emX}}

\begin{document}

\title{QaSAL: QoS-aware State-Augmented Learnable Algorithms for Coexistence of 5G NR-U/Wi-Fi \\
\thanks{This work was supported in part by the U.S. National Science Foundation under Grant 2034616.}
}

\author{
    \IEEEauthorblockN{Mohammad Reza Fasihi and
        Brian L. Mark}
    \IEEEauthorblockA{Dept. of Electrical and Computer Engineering
        and Wireless Cyber Center}
    \IEEEauthorblockA{George Mason University, Fairfax, Virginia, United States\\
        Email: mfasihi4@gmu.edu, bmark@gmu.edu}
}

\date{December 2024}

\maketitle

\begin{abstract}
With the increasing demand for wireless connectivity, ensuring the efficient coexistence of 
multiple radio access technologies in shared unlicensed spectrum has become an important issue. 
This paper focuses on optimizing Medium Access Control (MAC) parameters to enhance the coexistence of 5G New Radio in Unlicensed Spectrum (NR-U) and Wi-Fi networks operating in unlicensed spectrum with multiple
priority classes of traffic that may have varying quality-of-service (QoS) requirements. 
In this context, we tackle the coexistence parameter management problem by introducing a QoS-aware State-Augmented Learnable (QaSAL) framework, designed to improve network performance under various traffic conditions. Our approach augments the state representation with constraint information, enabling dynamic policy adjustments to enforce QoS requirements effectively. Simulation results validate the effectiveness of QaSAL in managing NR-U and Wi-Fi coexistence, demonstrating improved channel access fairness while satisfying a latency constraint for high-priority traffic.
\end{abstract}


\IEEEpeerreviewmaketitle

\section{Introduction}

\par The rapid growth in wireless connectivity demands, driven by diverse applications ranging from mobile communications to IoT networks, has led to an increasing reliance on unlicensed spectrum. These bands are shared by multiple Radio Access Technologies (RATs), such as 5G New Radio in Unlicensed Spectrum (NR-U) and Wi-Fi, introducing complex interference dynamics that can degrade network performance if not managed effectively. The primary challenge lies in achieving high network performance while maintaining fairness among different technologies sharing the same spectrum~\cite{Sathya:2020, Saha:2021, Hirzallah:2021}.

\par In this context, the coexistence of 5G NR-U and Wi-Fi networks introduces significant complexities. The two technologies use distinct channel access mechanisms: 5G NR-U relies on Listen Before Talk (LBT) procedures, while Wi-Fi employs Carrier Sense Multiple Access with Collision Avoidance (CSMA/CA). These mechanisms must operate harmoniously to minimize inter-network collisions and ensure equitable resource allocation. However, this is non-trivial due to the diverse Quality of Service (QoS) requirements, traffic patterns, and contention behaviors of the two networks. For instance, 5G NR-U often has stringent latency and reliability requirements, particularly for high-priority traffic, which can conflict with the opportunistic nature of Wi-Fi’s channel access mechanism.

\par Efficient coexistence is further complicated by the dynamic nature of unlicensed spectrum usage, where network configurations, traffic loads, and environmental factors fluctuate over time. These fluctuations demand adaptive and robust strategies to ensure that coexistence mechanisms can respond effectively to varying conditions~\cite{Muhammad:2020}. Without such strategies, the performance of both networks can degrade due to the increased collisions, higher latencies, and unfair resource allocation. For the coexistence of 5G NR-U and Wi-Fi networks on unlicensed spectrum, the problem managing contention and enhancing network performance can be formulated in terms of QoS-aware network utility maximization with respect to the Medium Access Control (MAC) parameters of both networks. One approach to this problem is based on regularized multi-objective reinforcement learning (RL), whereby the agent receives a reward representing a weighted combination of individual task-specific rewards~\cite{Mannor:2004, Moffaert:2013}. Although this method is widely used and can be effective, a significant drawback is the need for problem-specific selection of the weighting coefficients, which requires manual tuning and results in substantial computational overhead due to the extensive calibration process. Moreover, QoS constraints may be violated since they are not explicitly enforced in this approach.

\par Alternatively, QoS requirements can be explicitly defined within a constrained reinforcement learning (CRL) framework~\cite{Bhatnagar:2012}. In this setting, a single objective function referred to as the \textit{Lagrangian function} is maximized with respect to primal variables and minimized with respect to dual variables, each dual variable corresponding to a constraint in the original problem. A key advantage of this method is its ability to automatically adjust multiplier values, removing the necessity for manual tuning and thus reducing design complexity. By enforcing constraints in conjunction with RL algorithm, this approach can ensure that QoS constraints are met with high probability. Nonetheless, implementing this approach in dynamic environments, especially under varying network conditions, is challenging. Specifically, the penalty terms require extensive calibration and lack sufficient adaptability, ultimately leading to suboptimal performance in certain scenarios.

\par In this paper, we optimize the coexistence parameters of 5G NR-U and Wi-Fi networks to meet the QoS requirements of high-priority transmitters. We utilize dual variables in the CRL framework to track constraint violations over time~\cite{Fullana:2024, NaderiAlizadeh:2022, Uslu:2025}. The wireless network state is \textit{augmented} with these dual variables at each time step, serving as dynamic inputs to the learning algorithm. This augmentation improves the algorithm’s understanding of constraints and their environmental relationships, allowing the agent to adjust policies dynamically while minimizing reliance on indirect penalty mechanisms. A \textit{SimPy}-based simulation environment is implemented to model the MAC layer of 5G NR-U and Wi-Fi, supporting dynamic configurations for detailed performance evaluation. Transmitters from both networks operate in a saturated mode, representing high traffic conditions. This study builds upon our previous work~\cite{Fasihi:2024}, where we proposed a traffic priority-aware deep reinforcement learning (DRL) framework for dynamically adjusting contention window sizes to balance network performance and fairness. The current work expands on this by introducing a state-augmented learnable algorithm that directly integrates constraint variables into the state space. 

\par The remainder of the paper is organized as follows. In Section~\ref{sec:DRL-CPM}, we define our problem and study the DRL for CPM problem. Then, the primal-dual approach for our problem introduced in Section~\ref{sec:PrimalDual-CPM}. The QaSAL algorithm for solving the CPM problem is proposed in Section~\ref{sec:QaSALCPM}. Application of QaSAL framework to the coexistence of 5G NR-U and Wi-Fi is developed in~\ref{sec:QaSAL-CPM-Coexistence}. In Section~\ref{sec:simulationResults}, we present simulation results, which demonstrate that our proposed algorithm is able to protect the delay performance of high-priority traffic when contending with varying number of lower-priority traffic nodes for channel access. The paper is concluded in Section~\ref{sec:conclusion}.

\begin{figure*}
\centering
	\includegraphics[scale=0.33]{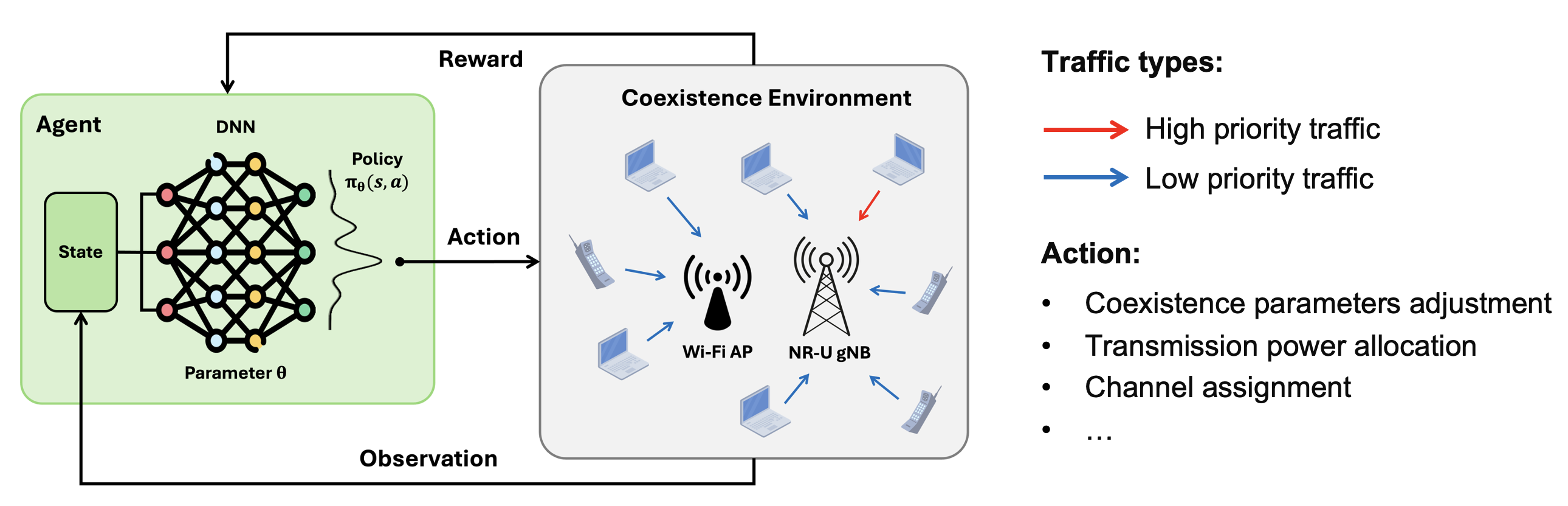}
    \caption{Parameterized optimization of CPM policy.}
	\label{fig:ML_Coexistence}
\end{figure*}

\section{Problem Formulation}
\label{sec:DRL-CPM}

\par Consider the coexistence of 5G NR-U and Wi-Fi in an unlicensed spectrum band. Let $\mathpzc{S}\subset\mathds{R}^n$ represent a compact set of \textit{coexistence environment states}, which captures the status of both the 5G NR-U and Wi-Fi networks. Given the state $\boldsymbol{S}_t\in\mathpzc{S}$, let $\mathbf{a}(\boldsymbol{S}_t)$ denote the vector of CPM decisions across both networks, where $\mathbf{a}:\mathpzc{S}\rightarrow\mathds{R}^a$ is the CPM function. The possible states and CPM decisions are described by a Markov Decision Process (MDP). The agent sequentially makes CPM decisions over discrete time steps $t\in\mathds{N}\cup\{0\}$ based on a policy $\pi$, leading to the performance vector $\boldsymbol{O}_t=\mathbf{f}(\boldsymbol{S}_t, \mathbf{a}(\boldsymbol{S}_t))\in\mathds{R}^m$ which captures various performance metrics of both networks, such as transmission delay, collision percentage, airtime efficiency, fairness, number of successful transmissions, channel utilization ratio, etc. Typically, in any MDP, the focus is on the accumulated performance over time, represented by the value function 
\begin{equation}
    \mathbf{V}_i(\pi)\triangleq\lim_{T\rightarrow\infty}\frac{1}{T}\mathds{E}_{\left(\boldsymbol{S},\mathbf{a}(\boldsymbol{S})\right)\sim\pi}\left[\sum_{t=0}^{T} f_i(\boldsymbol{S}_t, \mathbf{a}(\boldsymbol{S}_t))\right],
    \label{eq:ValueFunction}
\end{equation}
where $i=1,\ldots,m$. In practice, because the probability distribution of the environment is unknown, the objective is derived based on the \textit{ergodic average} network performance
\begin{equation}
    \Tilde{\mathbf{V}}_i(\pi)=\frac{1}{T}\sum_{t=0}^{T-1} f_i(\boldsymbol{S}_t, \mathbf{a}(\boldsymbol{S}_t)).
    \label{eq:ErgodicValueFunction}
\end{equation}
Note that the value functions $\Tilde{\mathbf{V}}_i(\pi)$, $i=1,...,m$, might be in conflict with each other, and a policy $\pi$ that is optimal for some $\Tilde{\mathbf{V}}_i$ may not be good for some other $\Tilde{\mathbf{V}}_j$. To include the QoS requirements in our CPM problem, let us define a concave utility
function $\mathpzc{U}:\mathds{R}^m\rightarrow\mathds{R}$ and a set of $c$ concave constraints defined in terms of a mapping $\boldsymbol{C}:\mathds{R}^m\rightarrow\mathds{R}^c$. The goal of the CPM problem is to specify the optimal vector of CPM decisions $\mathbf{a}(\boldsymbol{S}_t)$ for any given state $\boldsymbol{S}_t\in\mathpzc{S}$ that optimizes the utility function $\mathpzc{U}$ while ensuring the QoS requirements are satisfied.

\par The dynamic and complex nature of the coexistence environment as well as the different channel access mechanisms and the lack of coordination between two networks makes finding a proper solution for the CPM problem challenging. A learning-based optimization technique can be leveraged to maintain a fair and efficient coexistence. Therefore, we introduce a \textit{parameterized} CPM policy by replacing $\mathbf{a}(\boldsymbol{S})$ with $\mathbf{a}(\boldsymbol{S};\boldsymbol{\theta})$, where $\boldsymbol{\theta}\in\boldsymbol{\Theta}$ and $\boldsymbol{\Theta}$ denotes a finite-dimensional set of parameters (see Fig.~\ref{fig:ML_Coexistence}). Hence, the generic CPM problem can be defined as
\begin{subequations}
    \begin{align}
        \max_{\boldsymbol{\theta}\in\boldsymbol{\Theta}} \quad &
            \mathpzc{U}\left(\frac{1}{T}\sum_{t=0}^{T-1} \mathbf{f}(\boldsymbol{S}_t, \mathbf{a}(\boldsymbol{S}_t;\boldsymbol{\theta}))\right),
            \label{eq:ParameterizedCPM:Objective} \\
        \text{s.t.} \quad &
            \boldsymbol{C}\left(\frac{1}{T}\sum_{t=0}^{T-1} \mathbf{f}(\boldsymbol{S}_t, \mathbf{a}(\boldsymbol{S}_t;\boldsymbol{\theta}))\right) \geq \mathbf{0}
            \label{eq:ParameterizedCPM:Constraints}
    \end{align}
    \label{eq:ParameterizedCPM}
\end{subequations} 
where the maximization is performed over the set of parameters $\boldsymbol{\theta}\in\boldsymbol{\Theta}$. Note that the objective and constraints are derived based on the ergodic averages of the corresponding performance vectors. The goal of this paper is to develop a learning algorithm to solve (\ref{eq:ParameterizedCPM}) for any given coexistence environment state $\boldsymbol{S}_t\in\mathpzc{S}$.

\section{Gradient-Based CPM Algorithm In Dual Domain}
\label{sec:PrimalDual-CPM}

\par A customary approach to solve problem (\ref{eq:ParameterizedCPM}) is to consider a penalized version in the \textit{Lagrangian dual} domain. Formally, we introduce dual variables $\boldsymbol{\lambda}\in\mathds{R}_+^c$ associated with the constraints in (\ref{eq:ParameterizedCPM:Constraints}) and define the Lagrangian
\begin{multline}
    \mathpzc{L}_\pi(\boldsymbol{\lambda};\boldsymbol{\theta}) = \\ \mathpzc{U}\left(\frac{1}{T}\sum_{t=0}^{T-1} \mathbf{f}(\boldsymbol{S}_t, \mathbf{a}(\boldsymbol{S}_t;\boldsymbol{\theta}))\right) + \boldsymbol{\lambda}^T \boldsymbol{C}\left(\frac{1}{T}\sum_{t=0}^{T-1} \mathbf{f}(\boldsymbol{S}_t, \mathbf{a}(\boldsymbol{S}_t;\boldsymbol{\theta}))\right).
    \label{eq:ParameterizedLagrangian}
\end{multline}
The Lagrangian in (\ref{eq:ParameterizedLagrangian}) should be maximized over $\boldsymbol{\theta}$, while subsequently minimizing over the dual variables $\boldsymbol{\lambda}$, i.e.,
\begin{equation}
    \min_{\boldsymbol{\lambda}\in\mathds{R}_+^c} \quad \max_{\boldsymbol{\theta\in\Theta}} \quad \mathpzc{L}_\pi(\boldsymbol{\lambda};\boldsymbol{\theta}).
    \label{eq:MinMaxProblem}
\end{equation}
The advantage of replacing the objective in (\ref{eq:ParameterizedCPM}) with Lagrangian in (\ref{eq:ParameterizedLagrangian}) is that the latter can be optimized using any parameterized learning framework, such as standard reinforcement learning algorithms. One limitation of (\ref{eq:MinMaxProblem}) is the ambiguity in determining suitable values for the dual variables. The optimal choice for $\boldsymbol{\lambda}$ depends on the transition probability $p(\boldsymbol{S}_{t+1}|\boldsymbol{S}_t,\mathbf{a}(\boldsymbol{S}_t))$, which is typically unknown. This challenge can be addressed by \textit{dynamically adjusting} the $\boldsymbol{\lambda}$. To achieve this, we introduce an iteration index $k$, a step size $\boldsymbol{\eta}_k\in\mathds{R}_+^c$, and an epoch duration $T_0$ which is defined as the number of time steps between consecutive model parameter updates. Thus, the model parameters and dual variables are updated iteratively as
\begin{multline}
    \boldsymbol{\theta}_{k} = \arg\max_{\boldsymbol{\theta}\in\boldsymbol{\Theta}} \quad
    \mathpzc{U}\left(\frac{1}{T_0}\sum_{t=kT_0}^{(k+1)T_0-1} \mathbf{f}(\boldsymbol{S}_t, \mathbf{a}(\boldsymbol{S}_t;\boldsymbol{\theta}))\right) \\
    + \boldsymbol{\lambda}_k^T \boldsymbol{C}\left(\frac{1}{T_0}\sum_{t=kT_0}^{(k+1)T_0-1} \mathbf{f}(\boldsymbol{S}_t, \mathbf{a}(\boldsymbol{S}_t;\boldsymbol{\theta}))\right),
    \label{eq:LambdaUpdateIterationPrimalDual}
\end{multline}

\begin{equation}
    \boldsymbol{\lambda}_{k+1} = \left[\boldsymbol{\lambda}_{k}-\boldsymbol{\eta}_{\boldsymbol{\lambda}}\boldsymbol{C}\left(\frac{1}{T_0}\sum_{t=kT_0}^{(k+1)T_0-1} \mathbf{f}(\boldsymbol{S}_t, \mathbf{a}(\boldsymbol{S}_t;\boldsymbol{\theta}_k))\right)\right]^+, \quad \quad
    \label{eq:LambdaUpdateIteration}
\end{equation}
where $[x]^+:=\max(x,0)$. Note that the step size can be different for each QoS constraint. Dual variables $\boldsymbol{\lambda}$ at iteration $k$ are updated based on whether the constraints are violated or not. Dual variables are increased if the constraints are met and decreased if not, and the scale of update depends on the amount of violation at each time step.

\par The dual variable update algorithm in (\ref{eq:LambdaUpdateIterationPrimalDual}) and (\ref{eq:LambdaUpdateIteration}) presents a set of challenges that make it difficult to use in practice. Maximizing the Lagrangian in (\ref{eq:LambdaUpdateIterationPrimalDual}) requires knowledge of future network states, which is unattainable during execution, although it may be feasible during the training phase. Furthermore, achieving convergence to a feasible and near-optimal solution is only possible as the operation time $T$ approaches infinity. Finally, the optimal set of model parameters needs to be memorized for any given set of dual variables, which can be computationally expensive, especially during the execution phase. The \textit{state-augmentation} approach overcomes these limitations by embedding dual variables into the state space~\cite{Fullana:2024, NaderiAlizadeh:2022}. 

\begin{figure*}
\centering
\includegraphics[scale=0.32]{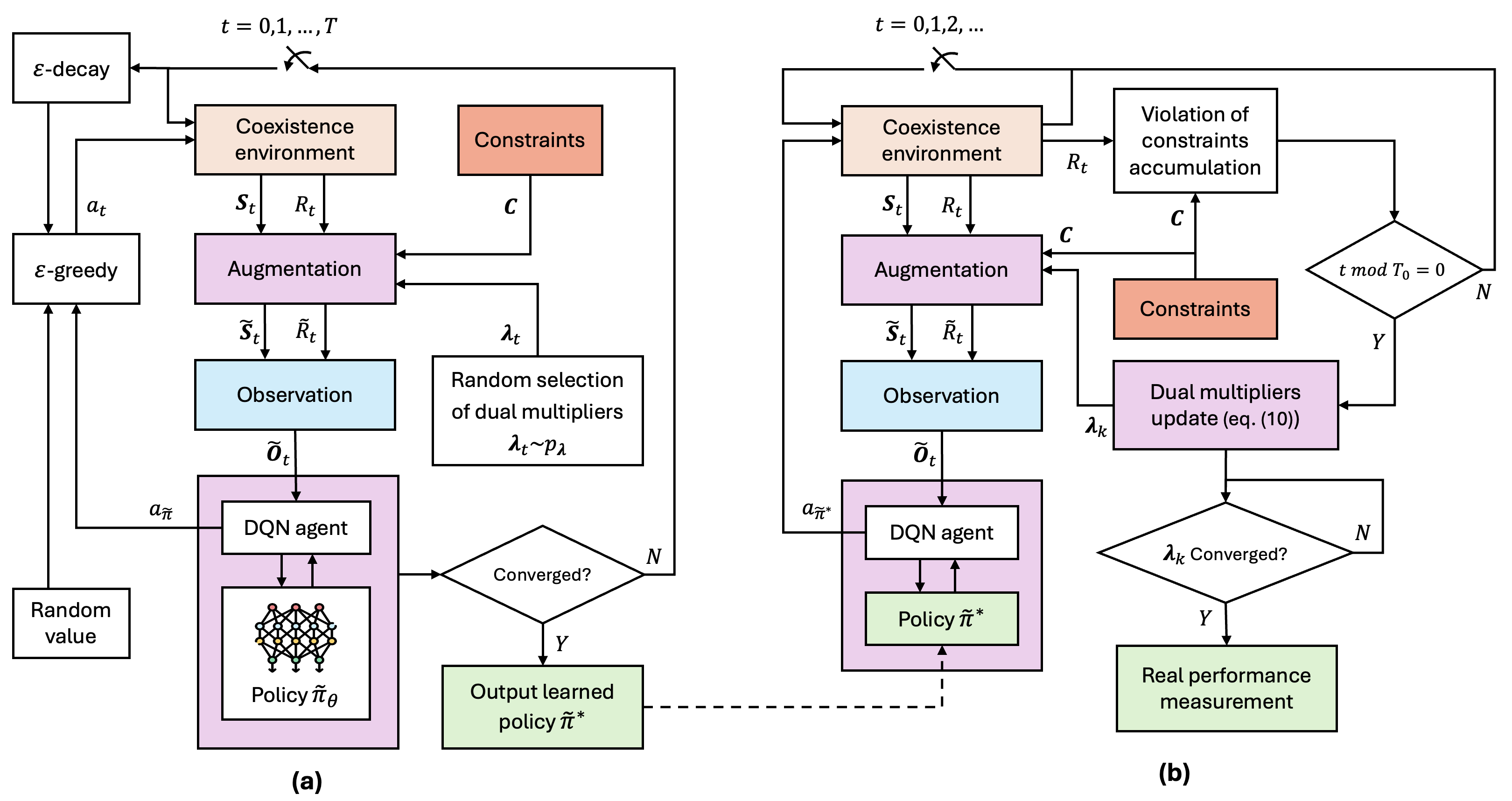}
    \caption{The flowchart of QaSAL algorithm for CPM: a) Training phase b) Execution phase.}
    \label{fig:QaSAL_Algorithm}
\end{figure*}

\section{Proposed QoS-aware State-Augmented Learnable Algorithm for CPM}
\label{sec:QaSALCPM}

In this section, we propose the QoS-aware State-Augmented Learnable Algorithm (QaSAL) for CPM problem. The key idea of state-augmentation is treating constraint satisfaction as a dynamic component of the agent's environment, which evolves in response to constraint violations through dual dynamics. By augmenting the state with dual variables, the agent learns policies that are constraint-aware and adaptable, leading to feasible, near-optimal solutions that traditional methods cannot guarantee.

\par We consider state $\boldsymbol{S}_t$ at time step $t$ of the $k$-th epoch. Augmentation of the dual variables $\boldsymbol{\lambda}_k$ into the state space results in a augmented state $\Tilde{\boldsymbol{S}}_t=(\boldsymbol{S}_t, \boldsymbol{\lambda}_k$). We introduce an alternative parameterization for the CPM policy, in which the CPM decisions $\boldsymbol{a}(\boldsymbol{S}, \boldsymbol{\theta})$ are represented via the parameterization $\boldsymbol{a}(\boldsymbol{\Tilde{S}}, \boldsymbol{\Tilde{\theta}})$, where $\boldsymbol{\Tilde{\theta}}\in\boldsymbol{\Tilde{\Theta}}$ denotes the set of parameters of the state-augmented CPM policy. Then, we define the augmented version of Lagrangian in (\ref{eq:ParameterizedLagrangian}) as
\begin{multline}
    \mathpzc{L}_\pi(\boldsymbol{\lambda};\boldsymbol{\Tilde{\theta}}) = \\ \mathpzc{U}\left(\frac{1}{T}\sum_{t=0}^{T-1} \mathbf{f}(\boldsymbol{\Tilde{S}}_t, \mathbf{a}(\boldsymbol{\Tilde{S}}_t;\boldsymbol{\Tilde{\theta}}))\right) + \boldsymbol{\lambda}^T \boldsymbol{C}\left(\frac{1}{T}\sum_{t=0}^{T-1} \mathbf{f}(\boldsymbol{\Tilde{S}}_t, \mathbf{a}(\boldsymbol{\Tilde{S}}_t;\boldsymbol{\Tilde{\theta}}))\right),
    \label{eq:ParameterizedAugmentedLagrangian}
\end{multline}
and formulate the augmented CPM policy optimization problem for any $\boldsymbol{\lambda}\sim p_{\boldsymbol{\lambda}}$ in (\ref{eq:MinMaxProblem}) as
\begin{equation}
    \boldsymbol{\Tilde{\theta}}^* = \arg\max_{\Tilde{\theta}\in\Tilde{\Theta}} ~\mathds{E}_{\boldsymbol{\lambda}\sim p_{\boldsymbol{\lambda}}}{\mathpzc{L}_\pi(\boldsymbol{\lambda};\boldsymbol{\Tilde{\theta}})}.
    \label{eq:MinMaxAugmentedProblem}
\end{equation}

\par Utilizing the augmented policy parameterized by (\ref{eq:MinMaxAugmentedProblem}), we substitute the dual variable update equation in (\ref{eq:LambdaUpdateIteration}) with the augmented version:
\begin{equation}
    \boldsymbol{\lambda}_{k+1} = \left[\boldsymbol{\lambda}_{k}-\boldsymbol{\eta}_{\boldsymbol{\lambda}}\boldsymbol{C}\left(\frac{1}{T_0}\sum_{t=kT_0}^{(k+1)T_0-1} \mathbf{f}(\boldsymbol{\Tilde{S}}_t, \mathbf{a}(\boldsymbol{\Tilde{S}}_t;\boldsymbol{\Tilde{\theta}}^*))\right)\right]^+.
    \label{eq:AugmentedLambdaUpdateIteration}
\end{equation}
Note that in the multiplier update equation (\ref{eq:AugmentedLambdaUpdateIteration}), the optimal parameters $\boldsymbol{\Tilde{\theta}}^*$ are utilized, which mitigates the challenge of storing the model parameters for any given set of dual variables in~(\ref{eq:LambdaUpdateIterationPrimalDual}). The training and execution procedures are summarized in Algorithms~\ref{alg:Algorithm1} and~\ref{alg:Algorithm2}, respectively. Fig.~\ref{fig:QaSAL_Algorithm} depicts the proposed QaSAL algorithm for CPM.

\section{QaSAL-CPM algorithm for coexistence of 5G NR-U and Wi-Fi on an Unlicensed Spectrum}
\label{sec:QaSAL-CPM-Coexistence}

\par We consider the coexistence of 5G NR-U gNBs and Wi-Fi APs on an unlicensed spectrum. Each network handles two types of transmitters: High-priority traffic transmitters with strict delay requirements, and low-priority traffic transmitters with less stringent delay constraints. NR-U gNB transmitters employ the listen-before-talk (LBT) mechanism, which is based on CSMA/CA with a binary exponential backoff procedure and supports four priority classes (PCs) that correspond to four access classes (ACs) in Wi-Fi's enhanced distributed channel access (EDCA)~\cite{ETSI_TS_137213}. Performing the LBT protocol can harm the latency performance of high-priority traffic such as Ultra-Reliable Low Latency Communication (URLLC) packets, especially when contending for channel access with low-priority traffic~\cite{Le:2021}. Additionally, the LBT protocol may cause additional transmission delay compared to the licensed spectrum due to the unpredictability of transmission opportunities. Therefore, the delay performance of high-priority traffic may not meet the requirements. 

\par In this section, we propose a QaSAL-CPM algorithm to protect the low-latency performance of high-priority transmitters while maintaining high fairness among two networks. We denote the high-priority and low-priority transmitters by \textit{PC1} (priority class 1) and \textit{PC3} (priority class 3), respectively. The 5G NR-U network's transmissions align with slot boundaries determined by the licensed spectrum numerology. If the channel is found idle, gNBs send reservation signals (RS) until the slot boundary to prevent other transmitters from occupying the channel. 

\begin{figure}
\begin{algorithm}[H]
    \caption{QaSAL algorithm's training phase for CPM.}
    \label{alg:Algorithm1}    
    \begin{algorithmic}[1]
        \STATE Sample $\boldsymbol{\Tilde{S}}=(\boldsymbol{S},\boldsymbol{\lambda})$ from augmented space $\mathpzc{S}\times\Lambda$.
        \STATE Calculate the augmented Lagrangian according to (\ref{eq:ParameterizedAugmentedLagrangian}).
        \STATE Use the parameterized learning algorithm to obtain policy according to (\ref{eq:MinMaxAugmentedProblem}): 
         $\boldsymbol{\Tilde{\theta}}^* = \arg\max_{\Tilde{\theta}\in\Tilde{\Theta}} {\mathpzc{L}_\pi(\boldsymbol{\lambda};\boldsymbol{\Tilde{\theta}})}$.
    \end{algorithmic}
    \textbf{Return:} Optimal model parameters $\boldsymbol{\Tilde{\theta}}^*$.
\end{algorithm}

\begin{algorithm}[H]
    \caption{QaSAL algorithm's execution phase for CPM.}
    \label{alg:Algorithm2}
    \textbf{Input:} Optimal model parameters $\boldsymbol{\Tilde{\theta}}^*$, step $\boldsymbol{\eta}_k$, QoS constraints $\boldsymbol{C}$, epoch $T_0$ \\
    \textbf{Initialize:} $\boldsymbol{\Tilde{S}}_0\leftarrow(\boldsymbol{S}_0, \boldsymbol{\lambda}_0=\boldsymbol{0})$
    \begin{algorithmic}[1]
    \STATE \textbf{for} $k=0,1,...$ \textbf{do}
    \STATE \quad Rollout $T_0$ steps with CPM decisions $\boldsymbol{a}(\boldsymbol{\Tilde{S}}, \boldsymbol{\Tilde{\theta}}^*)$.
    \STATE \quad Update dual dynamics according to (\ref{eq:AugmentedLambdaUpdateIteration}).
    \STATE \textbf{end for}
    \end{algorithmic}
    \textbf{Output:} Sequence of CPM decisions $\{\boldsymbol{a}_k; k=0,1,...\}$ 
\end{algorithm}
\end{figure}

\begin{figure*}
\begin{multicols}{3}
\centering
    \includegraphics[width=\linewidth]{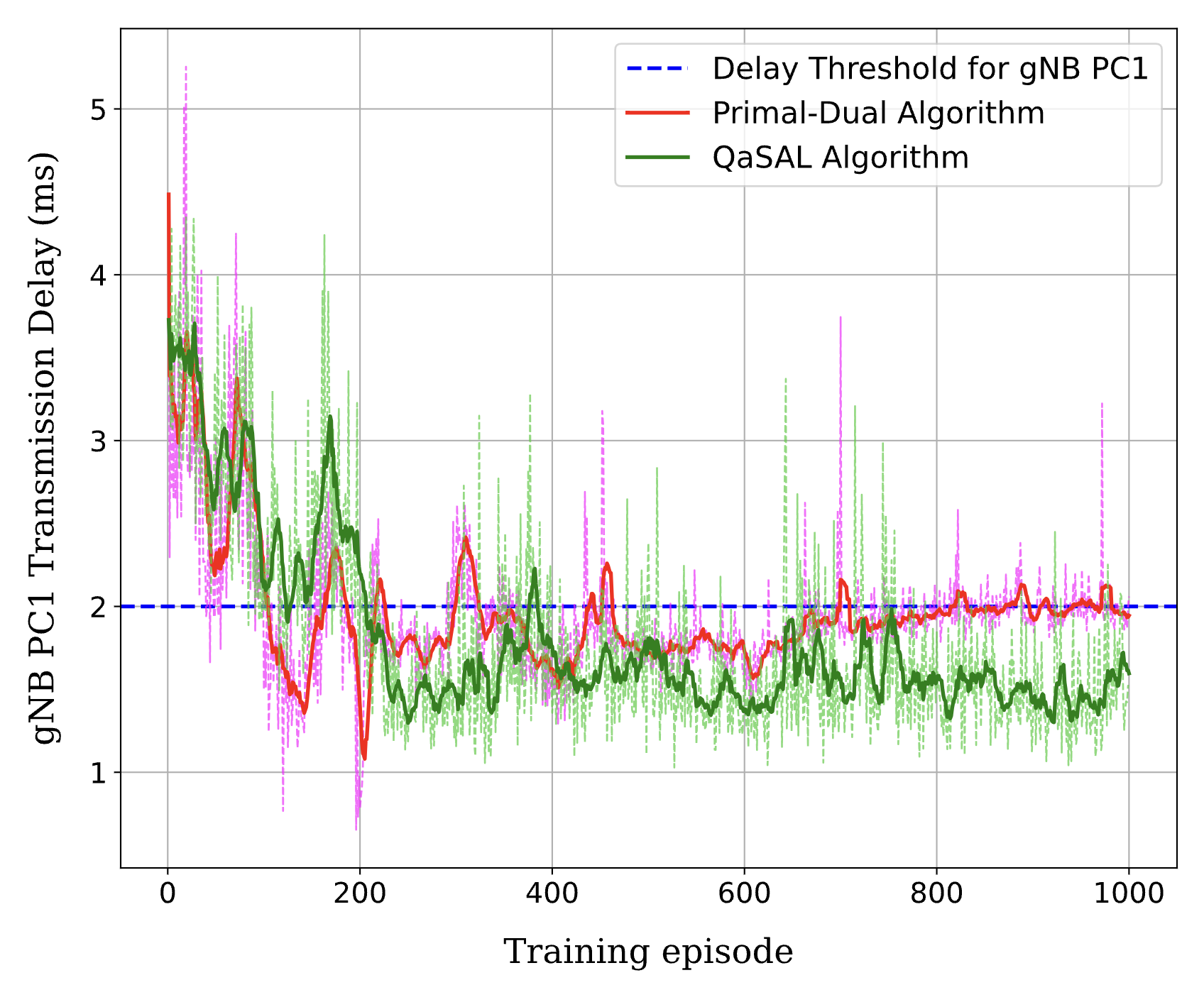}
        \caption{Training evolution of transmission delay of gNB PC1 transmitter ($D_{\text{th,PC1}}=2~\text{ms}$).}
        \label{fig:Training_Delay_PC1}
    \par
	\includegraphics[width=\linewidth]{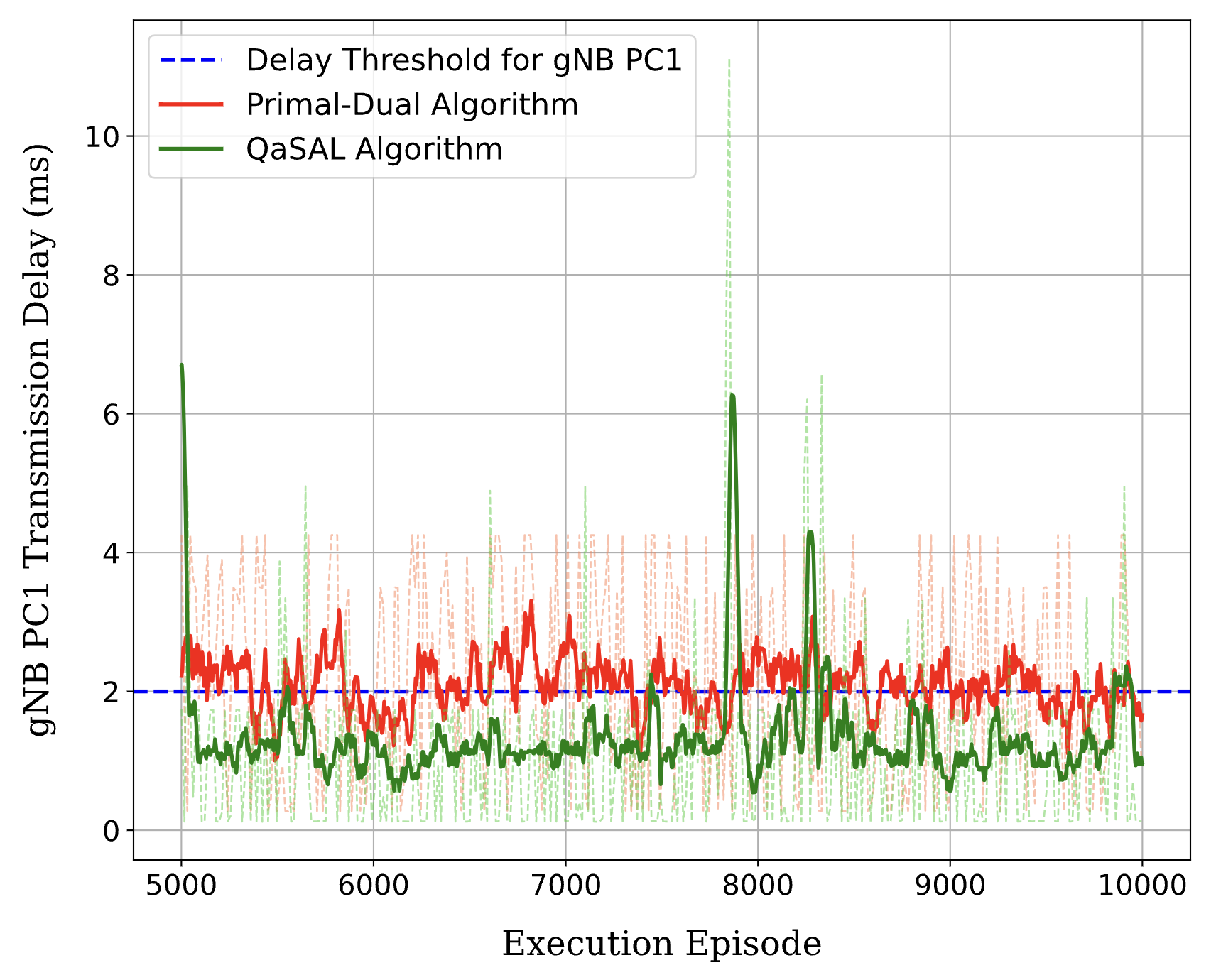}
        \caption{Execution evolution of transmission delay of gNB PC1 transmitter ($D_{\text{th,PC1}}=2~\text{ms}$).}
    	\label{fig:Execution_Delay_PC1}
    \par
    \includegraphics[width=\linewidth]{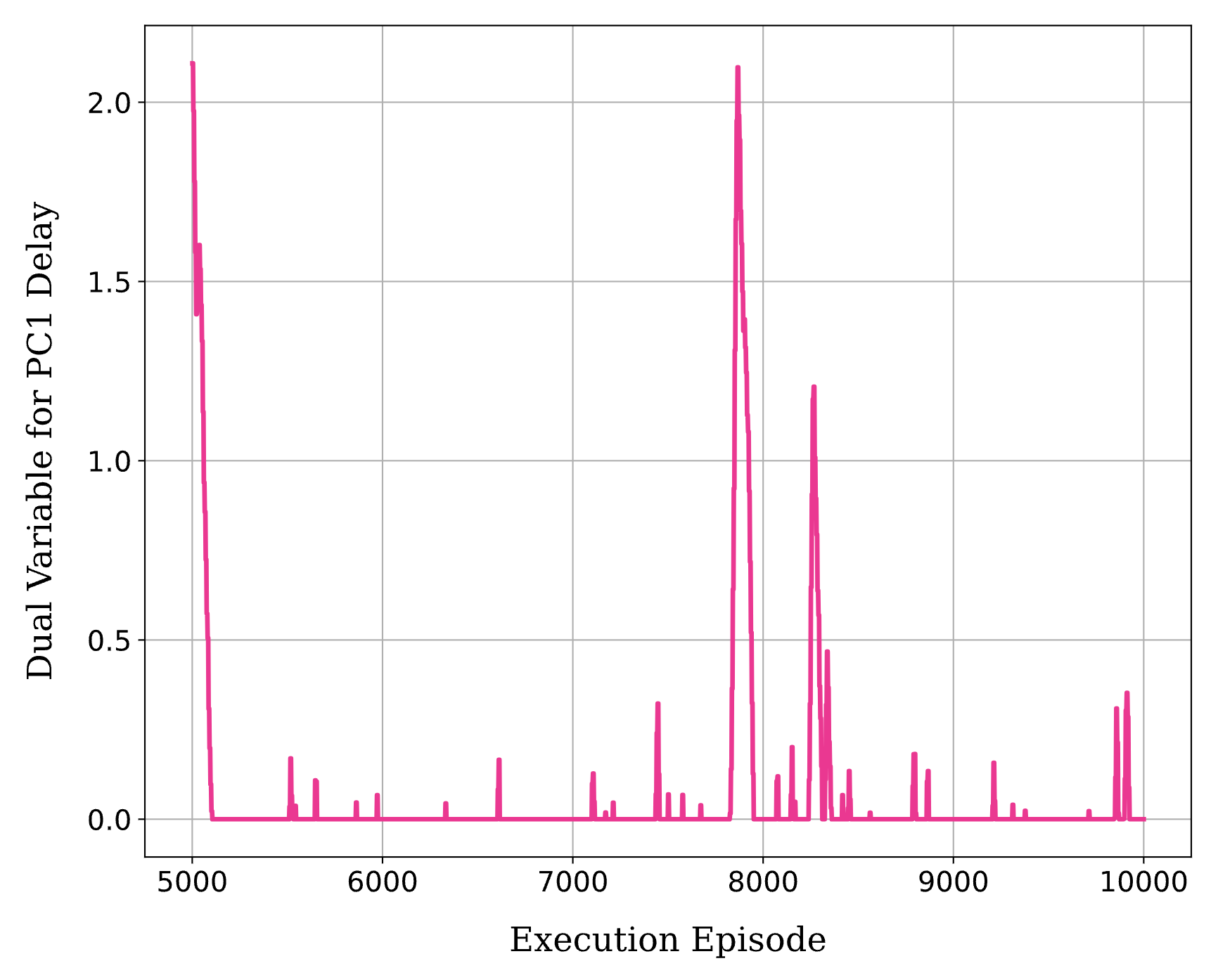}
        \caption{Dual variable evolution for delay constraint of gNB PC1 transmitter 
        ($D_{\text{th,PC1}}=2~\text{ms}$).}
    	\label{fig:Execution_Lambda_Delay_PC1}
\end{multicols}
\end{figure*}

\par The coexistence environment state at time step $t$, i.e., $\boldsymbol{S}_t$ is designed to comprehensively capture the dynamic environment of 5G NR-U and Wi-Fi coexistence on unlicensed spectrum. It encodes critical metrics that influence decision-making, including network performance and resource utilization such as the average and step-wise transmission delay of high-priority traffic, the percentage of collisions, channel airtime utilization, and Jain's Fairness Index (JFI). Additionally, it tracks trends in delay variation and short-term collision statistics to provide insights into ongoing network conditions.

\par The CPM decisions $\boldsymbol{a}(\boldsymbol{S}_t)$ is represented as a discrete selection from the set $\{0,1,\ldots,6\}$, where the maximum contention window (CW) size is calculated as $\text{CW}_{\text{max, PC}_i}=2^{\boldsymbol{a}_i+c}-1$, where $\boldsymbol{a}_i$ is the CPM decision for priority class $i$ and $c=0$ and $4$ for PC1 and PC3, respectively. This design directly influence the backoff timing, which affects each transmitter's access to the shared spectrum. Moreover, the performance function is designed as $f(\boldsymbol{S}_t, \mathbf{a}(\boldsymbol{S}_t))=\mathrm{(JFI)}_t$, where $\mathrm{(JFI)}_t$ denotes the JFI among both networks at time $t$. We define a constraint of the form $C(x) \leq 0$, where $C(x)=x-x_\text{th}$ with a threshold $x_{\text{th}}$. Consequently, the optimization problem in (\ref{eq:ParameterizedCPM}) can be formulated as
\begin{subequations}
    \begin{align}
        \max_{\pi} \quad &
            \frac{1}{T}\sum_{t=0}^{T-1} \mathrm{(JFI)}_t,
            \label{eq:ParameterizedCPMCoexistence:Objective} \\
        \text{s.t.} \quad &
            \frac{1}{T}\sum_{t=0}^{T-1} (\mathrm{D}_{\mathrm{PC}_1})_t\leq D_{\text{th}} 
            \label{eq:ParameterizedCPMCoexistence:Constraints}
    \end{align}
    \label{eq:ParameterizedCPMCoexistence}
\end{subequations}

\par We implement the QaSAL algorithm to solve the above CPM problem. The training phase involves the agent interacting with a simulated MAC layer environment, which models the behavior of 5G NR-U and Wi-Fi transmitters under realistic coexistence conditions. At each time step, the agent observes the current augmented state, selects an action which is adjusting contention window sizes, and observes performance functions. The constraint is carefully designed to penalize violations of QoS requirements. Experience replay is used to store and sample past transitions, and a target network is updated periodically to stabilize learning. Training is conducted across varying number of transmitters to ensure the generalization.

\par In the execution phase, the trained policy is deployed in the simulation environment to evaluate its effectiveness. The augmented state representation ensures that the agent adapts dynamically to changing network conditions, making real-time adjustments to contention window sizes to optimize system performance. The QaSAL algorithm effectively balances competing objectives, ensuring that high-priority traffic meets its delay requirements while promoting fairness between 5G NR-U and Wi-Fi networks. The hyperparameters of QaSAL algorithm are summarized in Table~\ref{DQN_param}. The step duration is selected to be large enough to include several transmission attempts to enable the accurate calculation of the transmission delay.

\begin{table}
    \centering
    \caption{Hyper-parameters of QaSAL algorithm}
    \begin{tabular}{| c | c |}
        \hline
        \textbf{Parameter} & \textbf{Value} \\ 
        \hline
        Interaction time $T$ & $20~\text{s}$ \\
        Step duration & $2.5~\text{ms}$ \\
        Discount factor & $0.99$ \\ 
        Replay buffer size & 100,000 \\
        Range of $\epsilon$ & 1 to 0.1 \\ 
        DQN learning rate & $10^{-4}$ \\
        Batch size & 16 \\
        Hidden layers & $32 \times 32 \times 32$\\
        $\eta_{\lambda}, \lambda_{\text{max}}, T_0$ in~(\ref{eq:AugmentedLambdaUpdateIteration}) & 0.1, 10, 5 \\
        \hline
    \end{tabular}
    \label{DQN_param}
\end{table}

\section{Simulation Results}
\label{sec:simulationResults}

\begin{figure*}
\begin{multicols}{2}
\centering
    \includegraphics[scale=0.25]{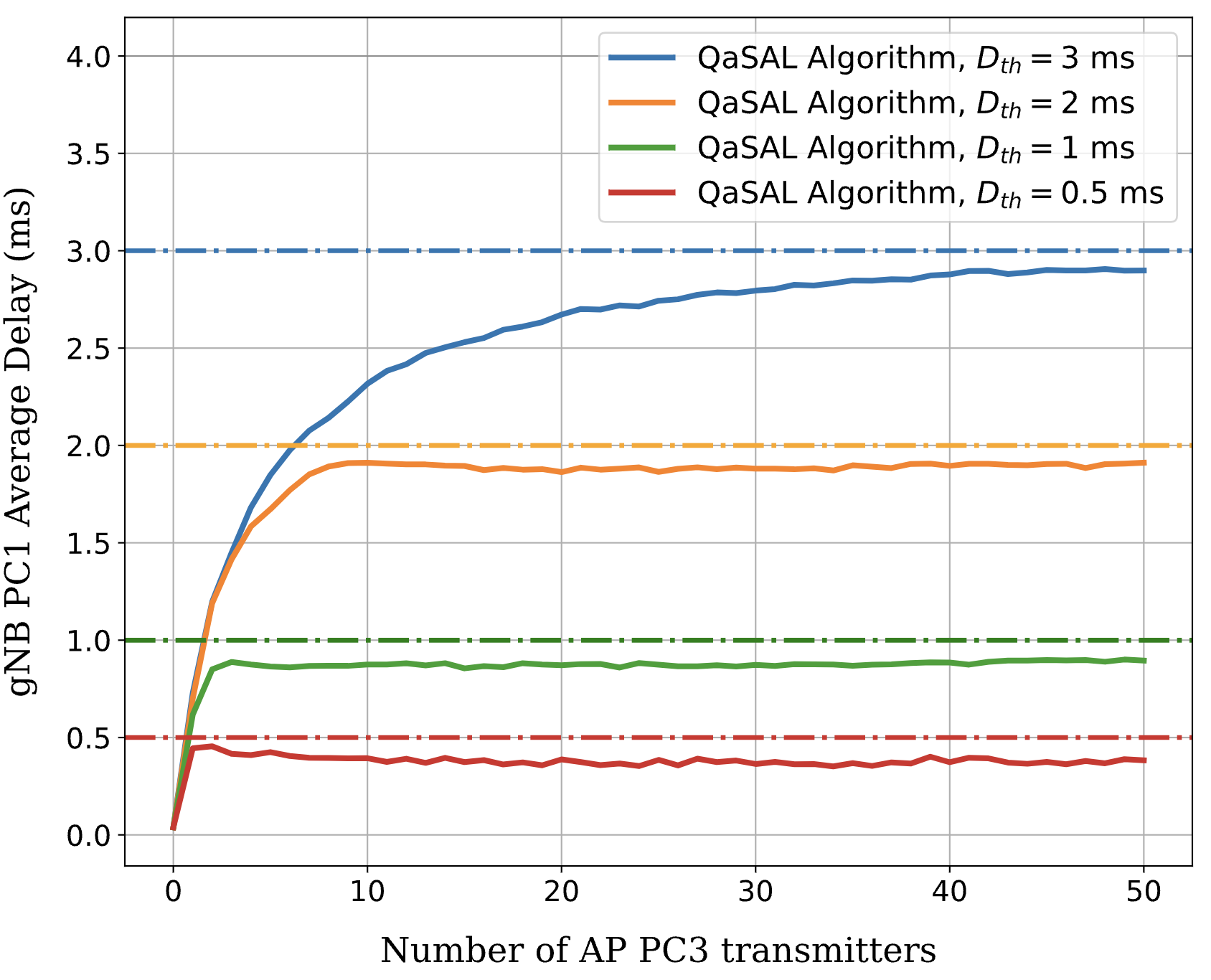}
        \caption{Average Transmission delay of PC1 transmitter with QaSAL algorithm for various delay thresholds and varying number of PC3 transmitters.}
        \label{fig:QaSAL_Delay}
    \par
	\includegraphics[scale=0.25]{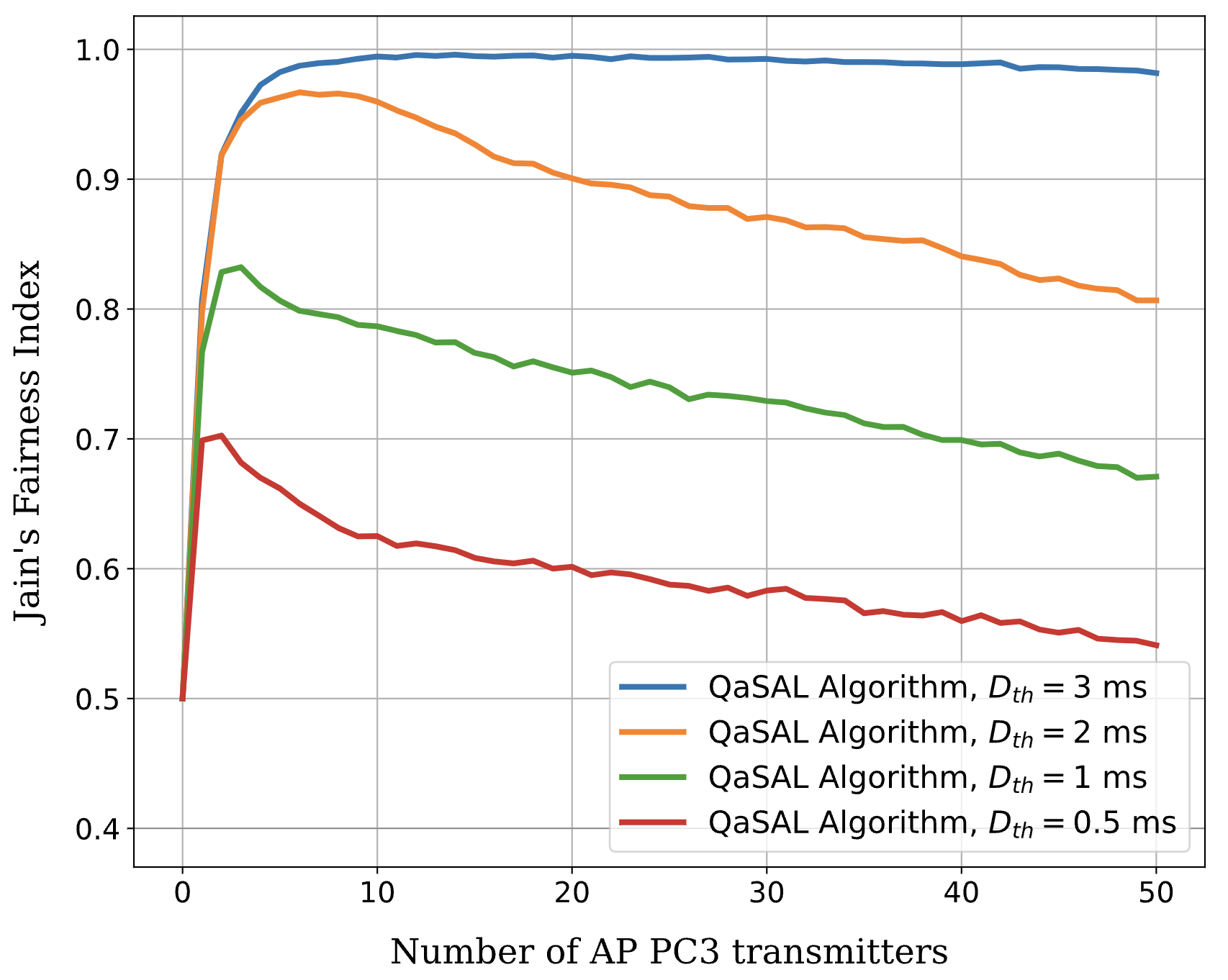}
    	\caption{JFI among Wi-Fi and NR-U networks with QaSAL algorithm for various delay thresholds and varying number of PC3 transmitters.}
    	\label{fig:QaSAL_JFI}
    \par
\end{multicols}
\end{figure*}

\par In this section, we highlight the performance of the proposed QaSAL algorithm. For the simulation scenario, we consider one gNB PC1 transmitter sharing the channel with varying number of AP PC3 transmitters, and evaluate the performance of primal-dual method (Section~\ref{sec:PrimalDual-CPM}) and QaSAL algorithm (Section~\ref{sec:QaSALCPM}). The simulations were conducted to analyze the algorithm's ability to balance fairness and delay metrics across different numbers of Wi-Fi's low-priority transmitters. 

\par Figs.~\ref{fig:Training_Delay_PC1} and \ref{fig:Execution_Delay_PC1} illustrate the transmission delay dynamics for gNB PC1 traffic coexisting with 25 AP PC3 transmitters during both the training and execution phases of the primal-dual and QaSAL algorithms, respectively. While the policy derived from the primal-dual algorithm can solve problem (\ref{eq:ParameterizedCPMCoexistence}) on average, it does not guarantee near-optimality at a specific episode $k$. In contrast, the QaSAL algorithm effectively enforces the constraint by incorporating dual variables into the state space, leading to improved constraint satisfaction. Fig.~\ref{fig:Execution_Lambda_Delay_PC1} illustrates the evolution of the dual variable associated with the delay constraint of gNB PC1, as defined in (\ref{eq:ParameterizedCPMCoexistence:Constraints}), during the execution phase. The dual variable adjusts dynamically in response to constraint violations, ensuring that the delay remains within the specified threshold.

\par Figs. \ref{fig:QaSAL_Delay} and \ref{fig:QaSAL_JFI} illustrate the average transmission delay for gNB PC1 traffic and JFI across both networks as the number of AP PC3 transmitters varies. The results demonstrate that the QaSAL algorithm effectively maintains the QoS requirements for high-priority traffic across different delay thresholds and varying number of transmitters while simultaneously optimizing fairness between the two networks.

\section{Conclusions}
\label{sec:conclusion}
In this paper, we introduced the QoS-aware State-Augmented Learnable (QaSAL) algorithm, a reinforcement learning-based approach designed to optimize the coexistence of 5G NR-U and Wi-Fi networks in unlicensed spectrum environments. By augmenting the network state with dual variables, our framework enables dynamic adaptation to QoS constraints, ensuring efficient spectrum sharing while maintaining low-latency transmission for high-priority traffic. Our results demonstrate that the QaSAL algorithm effectively balances network fairness with explicit delay constraints, outperforming traditional primal-dual methods in achieving constraint satisfaction. Unlike conventional approaches that require extensive parameter tuning, QaSAL learns optimal policies dynamically, improving system adaptability under varying traffic loads. 

\par Future work will focus on extending the QaSAL framework to multi-channel coexistence scenarios. Additionally, further improvements in training efficiency and multi-agent learning could enhance the scalability of QaSAL for next-generation wireless networks.

\bibliographystyle{IEEEtran}
\bibliography{IEEEabrv,main}

\end{document}